\documentclass[reprint,pre,superscriptaddress,amsmath,longbibliography]{revtex4-2}
\usepackage{graphicx,txfonts}
\usepackage[colorlinks,linkcolor=blue,citecolor=blue,anchorcolor=blue,urlcolor=blue]{hyperref}

\begin{document}

\title{Finite-size scaling of percolation on scale-free networks}

\author{Xuewei Zhao}
\affiliation{School of Physics, Hefei University of Technology, Hefei 230009, China}

\author{Liwenying Yang}
\affiliation{School of Physics, Hefei University of Technology, Hefei 230009, China}

\author{Dan Peng}
\affiliation{School of Journalism and Communication, Anhui University, Hefei 230601, China}

\author{Run-Ran Liu}
\affiliation{Alibaba Research Center for Complexity Sciences, Hangzhou Normal University, Hangzhou 311121, China}

\author{Ming Li}
\email{lim@hfut.edu.cn}
\affiliation{School of Physics, Hefei University of Technology, Hefei 230009, China}

\date{\today}

\begin{abstract}
Critical phenomena on scale-free networks with a degree distribution $p_k \sim k^{-\lambda}$ exhibit rich finite-size effects due to its structural heterogeneity. We systematically study the finite-size scaling of percolation and identify two distinct crossover routes to mean-field behavior: one controlled by the degree exponent $\lambda$, the other by the degree cutoff $K \sim V^{\kappa}$, where $V$ is the system size and $\kappa \in [0,1]$ is the cutoff exponent. Increasing $\lambda$ or decreasing $\kappa$ suppresses heterogeneity and drives the system toward mean-field behavior, with logarithmic corrections near the marginal case. These findings provide a unified picture of the crossover from heterogeneous to homogeneous criticality. In the crossover regime, we observe rich finite-size phenomena, including the transition from vanishing to divergent susceptibility, distinct exponents for the shift and fluctuation of pseudocritical points, and a numerical clarification of previous theoretical predictions.
\end{abstract}

\maketitle

\section{Introduction}

Due to the nontrivial topological features, network ensembles have emerged as a powerful platform for exploring the theories and methodologies of statistical physics~\cite{Dorogovtsev2008}. Unlike lattice systems, which possess homogeneous and spatially embedded topologies, networks often lack spatial constraints. Instead, network ensembles are typically characterized by their degree distribution $p_k$~\cite{Burda2001,Dorogovtsev2003,Park2004}, which gives the probability that a randomly selected node in the network has degree $k$, i.e., is connected to $k$ other nodes.

A typical network ensemble is the Erd\H{o}s-R\'{e}nyi (ER) random graph~\cite{Bollobas2001}, which follows a Poisson degree distribution. In the ER model, a fixed number $E$ of links is randomly placed among $V$ nodes, or alternatively, each pair of nodes is connected with a fixed probability. Owing to the absence of spatial constraints and the homogeneity of the degree distribution, critical phenomena in ER random graphs exhibit standard mean-field behavior~\cite{Dorogovtsev2008}. In contrast, scale-free (SF) networks, which serve as a key organizing principle in many real-world systems~\cite{Barabasi2009,Barabasi2016}, form another widely studied ensemble~\cite{Burda2001,Dorogovtsev2003,Goltsev2003,Park2004,Dorogovtsev2008}. This network ensemble is characterized by a power-law degree distribution
\begin{equation}
p_k \propto k^{-\lambda}, \quad k \geq m, \label{eq-pk}
\end{equation}
where $m$ is the minimum degree.

The primary difference between ER and SF networks lies in the tail of their degree distributions. In ER networks, node degrees are narrowly distributed around the mean. In contrast, the power-law degree distribution in SF networks allows the existence of high-degree nodes (hubs), introducing strong heterogeneity into the network structure. This heterogeneity profoundly affects both critical behavior and dynamical processes, with the exponent $\lambda$ playing a crucial role~\cite{Cohen2002,Goltsev2003,Lee2004,Castellano2006,Hong2007,Boccaletti2006,Barabasi2009,Barabasi2016}.

To understand how structural heterogeneity influences critical behavior, one of the most extensively studied models is percolation~\cite{Stauffer1991}, which provides a prototypical framework for exploring the emergence of large-scale connectivity in network systems~\cite{Newman2010,Li2021}. In the bond percolation model, each link is independently occupied with probability $P$. As $P$ increases, the system undergoes a percolation transition at a critical threshold $P_c$, above which a giant cluster emerges, containing a finite fraction of nodes connected by occupied bonds. This transition is accompanied by rich critical phenomena in the vicinity of $P_c$.

For SF networks, a well-known result is that for $2 < \lambda < 3$, the percolation threshold $P_c$ vanishes; that is, even an infinitesimally small occupation probability results in a giant cluster. Moreover, larger $\lambda$ values suppress the emergence of hubs, thereby reducing heterogeneity. When $\lambda > 4$, percolation on SF networks is generally believed to exhibit standard mean-field critical behavior, similar to ER networks~\cite{Cohen2002,Dorogovtsev2002,Leone2002}. However, for $\lambda<4$, various approaches have predicted different expressions for the critical exponents, particularly within the range $2<\lambda<3$. For instance, the Fisher exponent $\tau$ describing the cluster-size distribution has been reported as $(2\lambda - 3)/(\lambda - 2)$~\cite{Cohen2002} and $\lambda$~\cite{Lee2004,Kryven2017}, while the correlation-length exponent $\nu$ has been predicted as either $(\lambda - 1)/(3 - \lambda)$~\cite{Cohen2002} or $2/(3-\lambda)$~\cite{Radicchi}. More intriguingly, a susceptibility-like quantity defined as $\chi \equiv \sum_{k>1} C_k^2 / V$, where $C_k$ is the size of the $k$th largest cluster and the sum is over all clusters excluding the largest one, is suggested to vanish at criticality for $2 < \lambda < 3$~\cite{Cohen2002}. This contradicts the standard percolation behavior in which $\chi$ diverges at criticality, typically following finite-size scaling (FSS) as $\chi \sim V^{2d_f - 1}$, where $d_f$ is the volume fractal dimension of percolation clusters. A recent important advance in this direction is the work by Cirigliano et al.~\cite{Cirigliano2024}, which reconciles these differing predictions across the various $\lambda$ regimes and highlights that the FSS behavior can subtly depend on the detailed form of the degree distribution.

While much attention has been given to the role of the degree exponent $\lambda$, FSS behavior on SF networks is also strongly affected by the degree cutoff $K$, which limits the maximum node degree~\cite{Castellano2008}. The cutoff effectively tunes the heterogeneity of the network: smaller $\lambda$ increases heterogeneity by allowing more hubs, whereas smaller $K$ strictly restricts the presence of such hubs. Thus, the interplay between $\lambda$ and $K$ is expected to influence the structural heterogeneity of the network, which may, in turn, give rise to different FSS behaviors—though this relationship remains to be fully understood.

From extreme-value theory, even without an explicit upper bound $K$ in the degree distribution of Eq.~(\ref{eq-pk}), the finite system size $V$ naturally imposes a constraint on the maximum degree. This so-called \emph{natural cutoff} can be estimated such that the expected number of nodes with degree larger than $K$ remains finite as $V \to \infty$~\cite{Waclaw2008,Castellano2008,Li2021,Wang2025},
\begin{equation}
\int_K^V p_k\, dk \sim \mathcal{O}(1),
\end{equation}
yielding the scaling~\cite{Cohen2001},
\begin{equation}
K \sim V^{1/(\lambda - 1)}. \label{eq-K}
\end{equation}
This sets the intrinsic upper limit of the power-law degree distribution in Eq.~(\ref{eq-pk}). Even if a larger cutoff exponent is specified, the generated network will effectively revert to natural cutoff.

In real-world networks, however, degree cutoffs are often constrained by physical limits or design considerations, and are typically smaller than natural cutoff~\cite{Clauset2009}. Nonetheless, they usually grow with system size according to
\begin{equation}
K \sim V^{\kappa},
\end{equation}
where the cutoff exponent $\kappa$ ranges from $0$ to $1$. For $\kappa \to 1$, the cutoff significantly exceeds the average degree and allows for the presence of hubs; in contrast, when $\kappa \to 0$, the cutoff becomes finite and comparable to the average degree. A representative case is the \emph{structural cutoff} $\kappa = 1/2$, i.e., $K \sim \sqrt{V}$, which arises in uncorrelated SF networks that forbid multiple links~\cite{Chung2002,Burda2003,Boguna2004,Lee2006,Waclaw2008,Baek2012}. In such networks, the maximum degree should not exceed this structural limit.

In this paper, we employ the percolation model to demonstrate that the FSS behaviors on SF networks depends not only on the degree exponent $\lambda$ but also on the cutoff exponent $\kappa$. For $\lambda<4$, the FSS behavior varies continuously with $\kappa$: when $\kappa\to0$, standard mean-field behavior emerges regardless of $\lambda$. Thus, a crossover from the specialized critical behaviors of SF networks to the mean-filed behavior can be observed for any $\lambda$ by varying $\kappa$. In the regime $2<\lambda<3$, where the percolation threshold vanishes and conventional FSS becomes ill-defined, we show that proper FSS can still be observed near a dynamic pseudocritical point. In particular, the abnormal behaviors, such as the vanishing of susceptibility, can be mitigated by choosing small $\kappa$, thereby restoring a divergent susceptibility that satisfies the hyperscaling relation. For $\lambda>4$, the system is expected to exhibit mean-field criticality, with $\lambda$ and $\kappa$ affecting only the corrections to scaling.

The remainder of the paper is organized as follows. In Sec.~\ref{sec-ma}, we introduce the network and percolation models along with the simulation algorithms. In Sec.~\ref{sec-nr}, we present results on the FSS behavior with varying $\lambda$ and $\kappa$. Finally, a discussion is included in Sec.~\ref{sec-dis}.

\section{Model and algorithm}  \label{sec-ma}

\subsection{Scale-free network}

We consider percolation processes on SF networks generated by the configuration model~\cite{Newman2001,Molloy1995}. The network consists of $V$ nodes, where each node $i$ is assigned a degree $k_i$ drawn independently from the power-law distribution $p_k\sim k^{-\lambda}$. To implement the degree cutoff, we employ the inverse transform sampling method with an explicit upper bound $K = a V^{\kappa}$ on the degree~\cite{Newman1999}, where $a$ is $V$-independent parameter. The minimum degree is set as $m=2$ to ensure that the network is locally connected~\cite{Aiello2001}. In most cases, we set $a = 1$; however, for small values of $\kappa$, we increase $a$ to $2$ or $4$ so that the cutoff $K$ remains sufficiently larger than the minimum degree $m$. After assigning degrees to all sites, we construct the network by randomly pairing the resulting half-edges (stubs) to form undirected links, while rejecting self-loops and multiple links. This yields an ensemble of SF networks with the target degree distribution.

To maintain randomness and the absence of self-loops and multiple links, the degree cutoff cannot be larger than the structural cutoff, i.e., $\kappa \leq 1/2$~\cite{Boguna2004,Lee2006}. Consequently, the natural cutoff $V^{1/(\lambda-1)}$ cannot be achieved for $\lambda<3$. In addition, when $\kappa > 1/(\lambda-1)$, the degree cutoff always matches the natural cutoff, and $\kappa$ no longer affects the results. In other cases, the degree heterogeneity of the network is determined jointly by the exponents $\lambda$ and $\kappa$.

\subsection{Bond Percolation}

Standard bond percolation studies the formation of connected structures (clusters) in a network where each link is independently occupied with a given probability $P$. In this work, we adopt an equivalent dynamic link-insertion process to realize bond percolation. Specifically, starting from an empty network where all links are absent, one link is randomly selected and added at each discrete time step $T$, chosen from the set of links generated by the configuration model. For any time step $T$, the system corresponds to a bond percolation realization with an effective occupation probability $P = T / E$, where $E$ is the total number of links in the network.

During the link-insertion process, we monitor the size of the largest cluster $\mathcal{C}_1$ as a function of $T$, and define the one-step increment of $\mathcal{C}_1$ as~\cite{Manna2011,Fan2020}
\begin{equation}
\Delta \mathcal{C}_1(T)= \mathcal{C}_1(T+1) - \mathcal{C}_1(T).
\end{equation}
As the process proceeds, $\Delta \mathcal{C}_1$ initially grows as small clusters begin to merge. Near the percolation threshold $P_c$, the merging of large clusters leads to a sharp increase in $\mathcal{C}_1$, resulting in a peak in $\Delta \mathcal{C}_1$. After this peak, the growth of $\mathcal{C}_1$ slows down as most nodes become part of the largest cluster. Therefore, the fraction of inserted links $\mathcal{P}_V$, corresponding to the time step of the maximum $\Delta \mathcal{C}_1$, can be identified as a pseudocritical point of the system. Note that, $\mathcal{P}_V$ varies from realizations to realizations, termed as \emph{dynamic pseudocritical point}.

\subsection{Sampled quantities}

For each realization, we first generate a SF network with $V$ nodes. Then, we apply the link-insertion process and identify the dynamic pseudocritical point $\mathcal{P}_V$. Various quantities are sampled at $\mathcal{P}_V$~\cite{Shi2025}, including the size of the $k$th largest cluster $\mathcal{C}_k$ and the number $\mathcal{N}_s$ of clusters of size $s$. For a number of realizations, we calculate the following observables
\begin{itemize}
\item The mean pseudocritical point $P_V \equiv \langle \mathcal{P}_V \rangle$.
\item The standard deviation of the dynamic pseudocritical point $\sigma_V \equiv \sqrt{\langle \mathcal{P}_V^2 \rangle - \langle \mathcal{P}_V \rangle^2}$.
\item The mean size of the $k$th largest cluster $C_k \equiv \langle \mathcal{C}_k \rangle$.
\item A susceptibility-like quantity $\chi \equiv \langle\sum_{k>1} \mathcal{C}_k^2 \rangle/ V$, where the sum runs over all the clusters with the largest one excluded.
\item The cluster-number density $n(s,V) \equiv \langle \mathcal{N}_s\rangle/V$.
\end{itemize}
Here, $\langle \cdot \rangle$ denotes an ensemble average evaluated at $\mathcal{P}_V$, referred to as the event-based ensemble~\cite{Li2023,Li2024}.

\section{Numerical Results}  \label{sec-nr}

\subsection{$2<\lambda<3$}

\subsubsection{Dynamic pseudocritical point}

As shown by Molloy and Reed~\cite{Molloy1995,Molloy1998}, a giant component emerges in a network when the degree distribution satisfies $\langle k^2\rangle/\langle k\rangle > 2$. Accordingly, the percolation threshold on SF networks is given by~\cite{Newman2001}
\begin{equation}
P_c = \frac{\langle k\rangle}{\langle k(k-1)\rangle} = \frac{\zeta(\lambda-1,m)}{\zeta(\lambda-2,m) - \zeta(\lambda-1,m)}, \label{eq-pcd}
\end{equation}
where $\zeta(s,a) \equiv \sum_{n=0}^\infty (n+a)^{-s} = \sum_{n=a}^\infty n^{-s}$ is the Hurwitz zeta function, and $m$ is the minimum degree.

\begin{figure}
\centering
\includegraphics[width=\columnwidth]{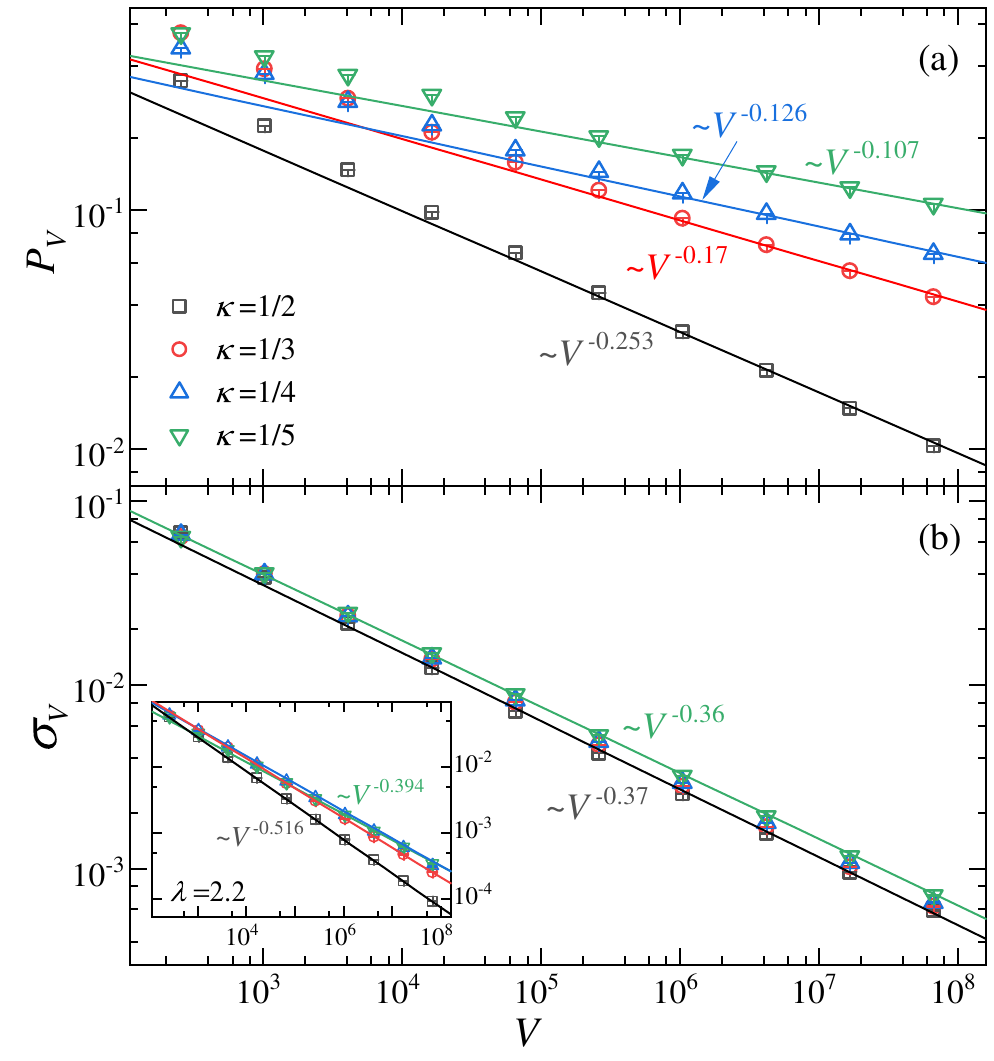}
\caption{(Color online) Asymptotic behaviors of dynamic pseudocritical points for $\lambda=2.5$. (a) The mean pseudocritical point $P_V$ is plotted against the system volume $V$ for different degree cutoffs, demonstrating a clear power-law scaling $P_V \sim V^{-1/\nu}$. This indicates that for any degree cutoff, the SF network exhibits a vanishing percolation threshold. The fit results further suggest that the exponent $1/\nu$ decreases as the degree cutoff decreases. (b) The fluctuation $\sigma_V$ of the dynamic pseudocritical point as a function of system volume $V$. The FSS of $\sigma_V \sim V^{-1/\nu'}$ implies a different exponent $\nu'$, which appears insensitive to the cutoff exponent $\kappa$. The lines indicate the fit result $1/\nu' = 0.36$ and $1/\nu' = 0.37$. The inset shows $\sigma_V \sim V^{-1/\nu'}$ for $\lambda=2.2$, suggesting the exponent $\nu'$ is still $\kappa$-dependent.}
\label{f1}
\end{figure}

For $2<\lambda<3$, the Hurwitz zeta function $\zeta(\lambda-2,m)$ diverges, resulting in a vanishing percolation threshold in the thermodynamic limit. Taking $\lambda = 2.5$ as an example, Fig.~\ref{f1} (a) shows the pseudocritical point $P_V$ as a function of system volume $V$ for various degree cutoffs. A clear FSS behavior $P_V \sim V^{-1/\nu}$ is observed for all cutoffs, confirming that $P_V\to P_c = 0$ as $V \to \infty$. Moreover, the exponent $1/\nu$ systematically decreases with decreasing cutoff, indicating that although hubs still promote global connectivity and ensure $P_c=0$, suppressing hub sizes amplifies finite-size effects and slows the convergence of $P_V$ toward zero.

To quantitatively understand this behavior, we adopt a truncated power-law degree distribution ($m \leq k \leq K$) in Eq.~(\ref{eq-pcd}). Using the asymptotic expansion of the Hurwitz zeta function~\cite{Bateman1953}, the critical point $P_c(K)$ for a given cutoff $K \sim V^{\kappa}$ behaves as
\begin{equation}  \label{eq-pcv}
P_c(K)-P_c\propto \left\{  \begin{array}{lr}
                              K^{3-\lambda}\sim V^{-\kappa(\lambda-3)}, & \lambda>3,    \\
                              (\ln K)^{-1} \sim (\ln V)^{-1},           & \lambda=3,    \\
                              K^{\lambda-3}\sim V^{-\kappa(3-\lambda)}, & 2<\lambda<3.
                           \end{array} \right.
\end{equation}
For natural cutoff, $\kappa = 1/(\lambda - 1)$, Eq.~(\ref{eq-pcv}) recovers the result of Ref.~\cite{Cohen2002}. For $2<\lambda<3$, the natural cutoff exceeds the structural cutoff $V^{1/2}$, and thus is not feasible for random SF networks without multiple links.

\begin{table}
\caption{The fit results of the critical exponents for percolation on SF networks. The exponents $1/\nu'$ and $d_f$ are extracted by fitting $\sigma_V$ and $C_1$ to the FSS forms in Eqs.~(\ref{eq-fspv}) and (\ref{eq-c1}), respectively. For $\lambda = 4.5$, a logarithmic correction is included via Eq.~(\ref{eq-c1ln}) to obtain $d_f$. The percolation threshold $P_c$ and exponent $1/\nu$ are obtained by fitting $P_V$ to Eq.~(\ref{eq-fpv}). For $2 < \lambda < 3$, the threshold vanishes ($P_c = 0$). For $\lambda = 3.3$, $3.5$, $3.8$, $4.5$, and $6$, the thresholds given by Eq.~(\ref{eq-pcd}) are $0.1730$, $0.2687$, $0.3889$, $0.5901$, and $0.8135$, respectively. Missing entries indicate that stable fitting was not achievable under the corresponding parameters. Here, NC is an abbreviation for natural cutoff.}  \label{t1}
\begin{ruledtabular}
\begin{tabular}{llllll}
\text{ $\lambda$}   &  \multicolumn{1}{c}{$\kappa$}  &  \multicolumn{1}{c}{$d_{f}$} &  \multicolumn{1}{c}{$1/\nu$} & \multicolumn{1}{c}{$1/\nu'$} & \multicolumn{1}{c}{$P_c$}  \\
\hline
2.2  &  \multicolumn{1}{c}{1/2}  & 0.39(1)    & 0.39(1)  & 0.516(8)  &  0.00001(3) \\
     &  \multicolumn{1}{c}{1/3}  & 0.500(5)   & 0.24(1)  & 0.448(4)  &  0.000(3)   \\
     &  \multicolumn{1}{c}{1/4}  & 0.555(5)   & 0.18(1)  & 0.417(5)  &  0.005(5)   \\
     &  \multicolumn{1}{c}{1/5}  & 0.588(6)   & 0.14(1)  & 0.394(6)  &  0.003(2)   \\
2.5  &  \multicolumn{1}{c}{1/2}  & 0.413(7)   & 0.253(2) & 0.37(1)   &  0.00002(5) \\
     &  \multicolumn{1}{c}{1/3}  & 0.511(6)   & 0.17(1)  & 0.36(1)   &  -0.001(2)  \\
     &  \multicolumn{1}{c}{1/4}  & 0.558(6)   & 0.126(7) & 0.36(1)   &  -0.004(4)  \\
     &  \multicolumn{1}{c}{1/5}  & 0.585(4)   & 0.107(5) & 0.36(1)   &  0.000(9)   \\
     &  \multicolumn{1}{c}{1/6}  & 0.60(1)    & 0.086(6) & 0.36(1)   &  0.00(1)   \\
2.8  &  \multicolumn{1}{c}{1/2}  & 0.462(5)   & 0.11(2)  & 0.17(2)   & -0.002(3)  \\
     &  \multicolumn{1}{c}{1/3}  & 0.533(5)   & 0.07(3)  & 0.28(1)   &  0.00(2)   \\
     &  \multicolumn{1}{c}{1/4}  & 0.57(1)    & 0.07(3)  & 0.32(1)   &  0.02(2)   \\
     &  \multicolumn{1}{c}{1/5}  & 0.588(4)   & --       & 0.32(1)   &  --    \\
     &  \multicolumn{1}{c}{1/6}  & 0.60(1)    & --       & 0.33(1)   &  --   \\
3.3  &  \multicolumn{1}{c}{NC}   & 0.564(4)   & 0.14(2)  & 0.18(2)   & 0.173(3)   \\
3.5  &  \multicolumn{1}{c}{NC}   & 0.60(1)    & 0.19(2)  & 0.21(1)   & 0.271(3)   \\
     &  \multicolumn{1}{c}{1/3}  & 0.60(1)    & 0.17(1)  & 0.24(1)   & 0.269(2)   \\
     &  \multicolumn{1}{c}{1/4}  & 0.61(1)    & 0.14(3)  & 0.27(1)   & 0.269(6)   \\
     &  \multicolumn{1}{c}{1/5}  & 0.618(7)   & 0.11(1)  & 0.28(1)   & 0.26(1)    \\
     &  \multicolumn{1}{c}{1/6}  & 0.62(1)    & --       &  0.29(1)  &  --  \\
     &  \multicolumn{1}{c}{1/8}  & 0.633(4)   & --       &  0.30(1)  & --   \\
3.8  &  \multicolumn{1}{c}{NC}   & 0.620(8)   & 0.27(1)  & 0.26(2)   & 0.391(1)   \\
4.5  &  \multicolumn{1}{c}{NC}   & 0.67(2)    & 0.32(1)  & 0.33(1)   & 0.5908(1)\\
     &  \multicolumn{1}{c}{1/4}  & 0.668(4)   & 0.33(1)  & 0.33(2)   & 0.590(1)   \\
     &  \multicolumn{1}{c}{1/5}  & 0.667(6)   & 0.33(1)  & 0.33(2)   & 0.5906(3) \\
6    &  \multicolumn{1}{c}{NC}   & 0.666(2)   & 0.333(5) & 0.33(1)   & 0.8135(1)   \\
     &  \multicolumn{1}{c}{1/6}  & 0.666(1)   & 0.33(1)  & 0.326(5)  & 0.8135(2)
\end{tabular}
\end{ruledtabular}
\end{table}

Equation (\ref{eq-pcv}) thus explains the numerical results in Fig.~\ref{f1} (a): the scaling exponent $1/\nu = \kappa(3 - \lambda)$ decreases as $\kappa$ decreases. To extract $\nu$ quantitatively, we perform least-squares fits of $P_V$ as a function of $V$, applying a lower cutoff $V_{\min}$ to evaluate the stability of fit. Accordingly, we adopt the following FSS ansatz
\begin{equation}
P_V = P_c + V^{-1/\nu}(a_0 + a_1 V^{-\omega}), \label{eq-fpv}
\end{equation}
where $\omega$ is the leading correction-to-scaling exponent. In general, we select the smallest $V_{\min}$ for which the reduced chi-square is close to unity and remains stable when increasing $V_{\min}$.

We note that quenched randomness and structural heterogeneity introduce strong finite-size corrections~\cite{Wu2007}, which often prevent stable fits when $\omega$ is treated as a free parameter in Eq.~(\ref{eq-fpv}). In such cases, we instead attempt a series of fits using fixed values of $\omega$ to identify a stable fitting result. However, due to the lack of a theoretical basis for selecting $\omega$, the final estimates represent an overall consideration of these fits, limiting the precision of the extracted exponents. In some instances, the fitting results vary significantly with different fixed $\omega$, so that no stable and consistent estimate can be obtained with the current data and the fit function. Nevertheless, this limitation may be overcome in future studies with access to larger-scale simulations, which are beyond our current computational capability.

The fitted values of $1/\nu$ are summarized in Table~\ref{t1}. For $2 < \lambda < 3$, we present results for $\lambda = 2.2$, $2.5$, and $2.8$, where the extracted values of $1/\nu$ agree with the theoretical prediction $1/\nu = \kappa(3 - \lambda)$ within one or two error bars. As $\lambda$ approaches $3$, the exponent $1/\nu = \kappa(3 - \lambda)$ becomes increasingly small, and Eq.~(\ref{eq-pcv}) suggests a logarithmic asymptotic behavior. In such cases, a stable fit using Eq.~(\ref{eq-fpv}) becomes more difficult. Indeed, for $\lambda = 2.8$, the fitted values of $1/\nu$ are systematically larger than the predicted ones, and for small $\kappa$, we are unable to obtain stable and consistent fits due to strong finite-size corrections and the limited range of system sizes accessible in our simulations. Nevertheless, the overall trend that $1/\nu$ decreases with decreasing $\kappa$ remains robust.

The percolation threshold $P_c$ is also estimated in the fitting procedure (see Table~\ref{t1}). In the range $2 < \lambda < 3$, the estimated $P_c$ is consistent with zero within at most two error bars. These results suggest that a vanishing threshold remains compatible with the data, although strong finite-size effects may obscure the asymptotic behavior.

In the limit $\kappa \to 0$, where the degree cutoff becomes independent of system size, the SF network effectively reduces to a configuration model with a fixed maximum degree. In this regime, the system is expected to exhibit mean-field percolation behavior with $\nu = 3$. However, within the range of $\kappa$ accessible to our simulations, the fitted values of $1/\nu$ continue to decrease monotonically with decreasing $\kappa$, and can even fall below the mean-field value $1/3$. This indicates that $1/\nu$ may exhibit a sharp crossover from a vanishing value predicted by Eq.~(\ref{eq-pcv}) to the mean-field value, as $\kappa \to 0$, rather than varying smoothly. Whether this crossover is truly discontinuous or smoothed out in the infinite-size limit remains an open question, requiring simulations on larger systems beyond current computational capabilities.

To further characterize the asymptotic behavior of dynamic pseudocritical points, we examine the standard deviation $\sigma_V$ of $\mathcal{P}_V$. Figure~\ref{f1} (b) shows $\sigma_V$ versus $V$, revealing an FSS behavior of the form
\begin{equation}
\sigma_V \sim V^{-1/\nu'}.
\end{equation}
For most of the percolation systems, it is known that $\nu' = \nu$~\cite{Li2024a}, implying that the pseudocritical shift and its fluctuations vanish at the same rate. However, fitting $\sigma_V$ using the FSS ansatz
\begin{equation}
\sigma_V = V^{-1/\nu'}(a_0 + a_1 V^{-\omega}), \label{eq-fspv}
\end{equation}
reveals that $\nu'$ is systematically smaller than $\nu$, i.e., $\nu' < \nu$.

\begin{figure}
\centering
\includegraphics[width=\columnwidth]{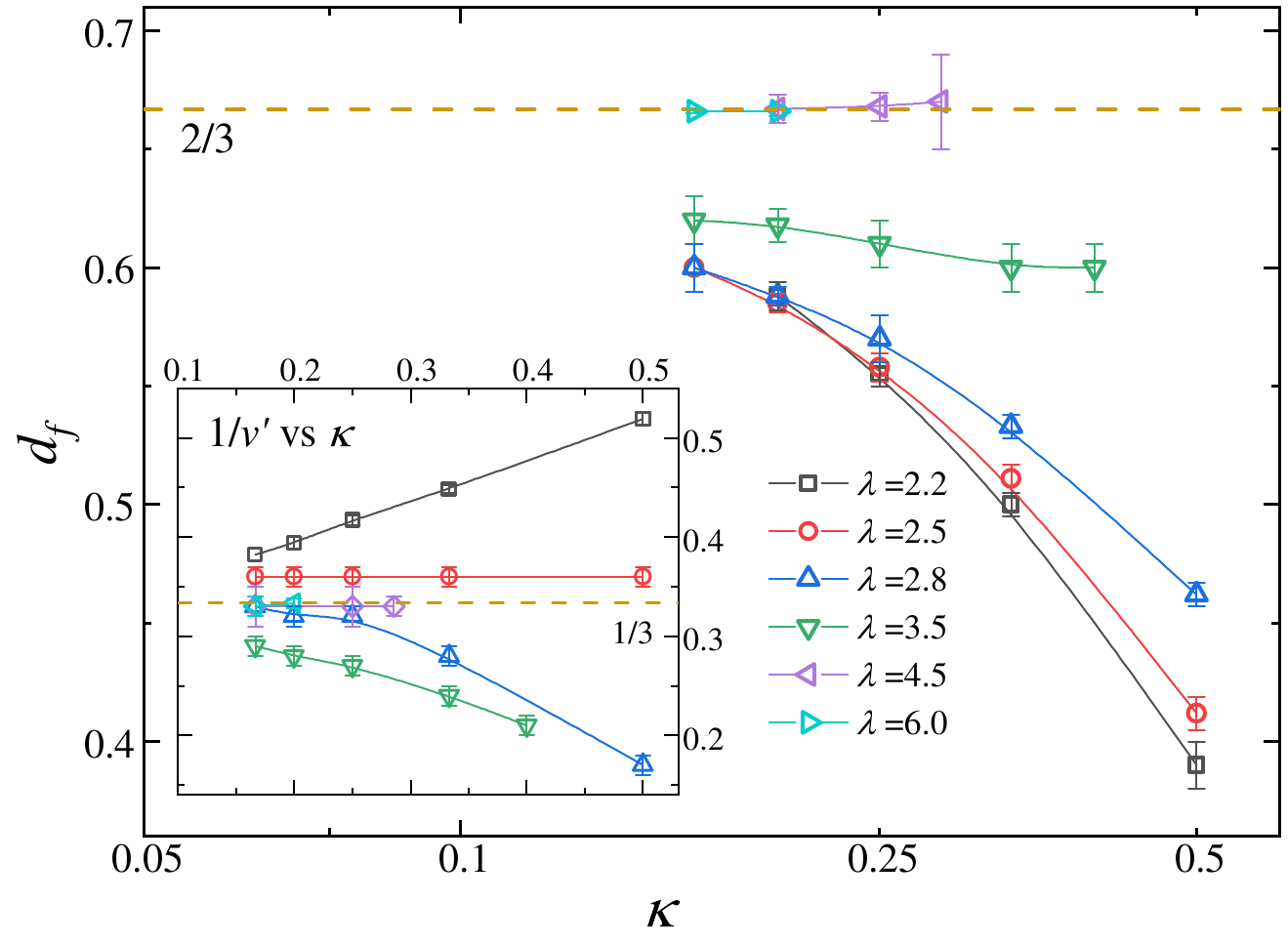}
\caption{(Color online) The fitted fractal dimension $d_f$ as a function of the cutoff exponent $\kappa$ for different $\lambda$. The scatters represent the results in Table~\ref{t1}, and golden dashed line indicates the mean-field value $d_f=2/3$. The inset shows the exponent $1/\nu'$ in Table~\ref{t1} as a function of $\kappa$, where the golden dashed line is also for the mean-field value $\nu'=3$. The plot demonstrate the trend that both $d_f$ and $\nu'$ converge towards the mean-field values for $\kappa\to0$.}
\label{f2}
\end{figure}

More intriguingly, in contrast to $1/\nu$, which systematically decreases with decreasing degree cutoff, the exponent $1/\nu'$ shows a nontrivial and $\lambda$-dependent trend: it increases for $\lambda = 2.8$, decreases for $\lambda = 2.2$, and remains nearly unchanged for $\lambda = 2.5$, as shown in Table~\ref{t1} and Fig.~\ref{f1} (b). This seemingly irregular behavior can be understood as a convergence toward the mean-field value. In the limit $\kappa \to 0$, the percolation on SF networks should follow the mean-field behavior with $\nu'=\nu=3$. Therefore, as the cutoff decreases, $1/\nu'$ approaches the mean-field value $1/3$ from above or below, depending on its initial deviation -- decreasing if initially larger than $1/3$, increasing if smaller. This convergence is clearly illustrated in the inset of Fig.~\ref{f2}, which shows $1/\nu'$ as a function of $\kappa$.

Furthermore, similar discrepancies between $\nu'$ and $\nu$ have been observed in high-dimensional percolation with open boundaries~\cite{Li2024a}. In this case, it has been argued that $\nu'$, rather than $\nu$, captures the true correlation-length scaling. Accordingly, we argue that for SF networks as well, $\nu$ does not represent the correlation-length exponent; instead, $\nu'$ should be taken as the correct one. More evidences for this argument will be demonstrated in the following sections.

\subsubsection{Fractal dimension}

\begin{figure}
\centering
\includegraphics[width=\columnwidth]{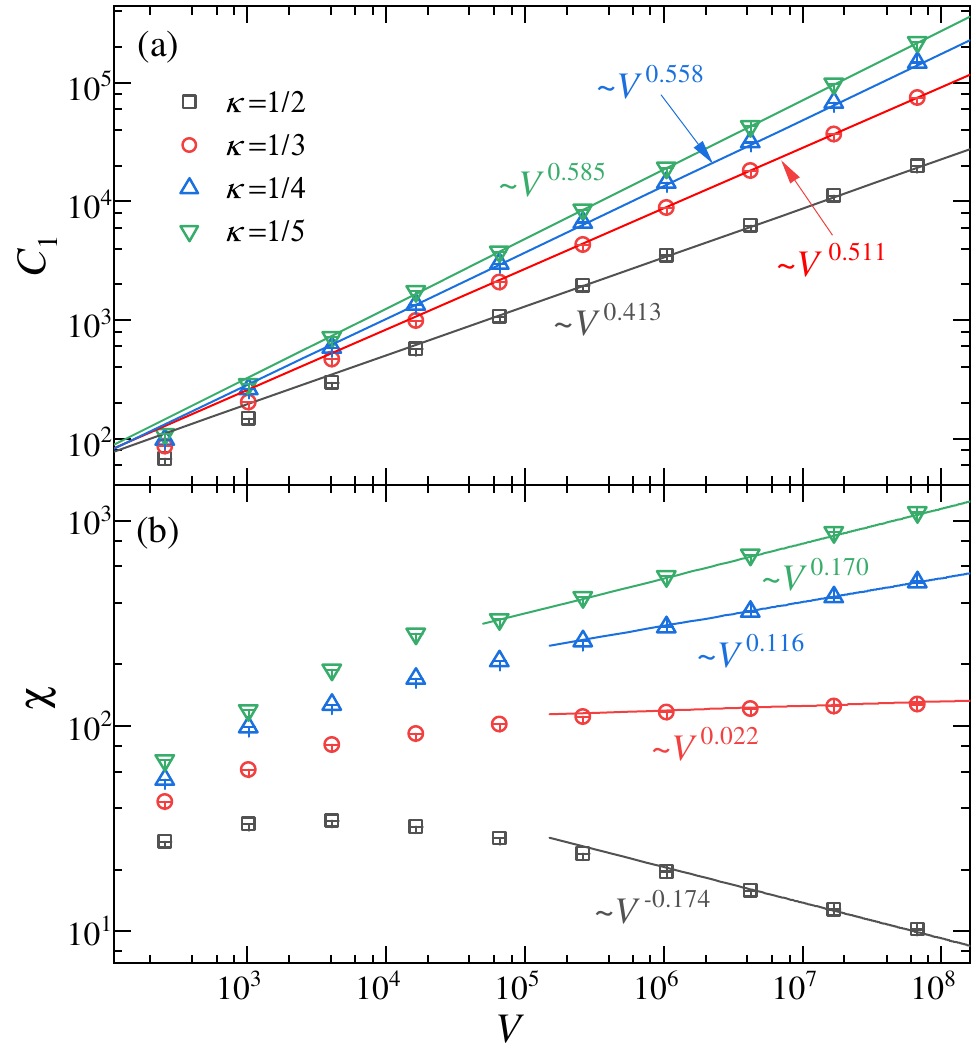}
\caption{(Color online) The FSS behaviors of the size of the largest cluster $C_1$ and the susceptibility-like quantity $\chi$ for $\lambda=2.5$. (a) $C_1$ as a function of the system volume $V$ for different degree cutoffs. The lines indicate the scaling $C_1 \sim V^{d_f}$, where $d_f$ is the fitted fractal dimension listed in Table~\ref{t1}. (b) $\chi$ as a function of $V$ for different degree cutoffs. The lines indicate the scaling $\chi \sim V^{2d_f - 1}$, with $d_f$ from Table~\ref{t1}.}
\label{f3}
\end{figure}

For $2<\lambda<3$, the percolation threshold $P_c$ vanishes, and hence at $P_c = 0$, only isolated sites exist regardless of system size. As a result, FSS behaviors cannot be observed at $P_c$. To overcome this difficulty, we investigate the FSS behaviors at the dynamic pseudocritical point $\mathcal{P}_V$, i.e., within the event-based ensemble.

Figure~\ref{f3} (a) shows the mean size of the largest cluster $C_1$, sampled at $\mathcal{P}_V$, as a function of $V$ for $\lambda = 2.5$. A clear power-law scaling $C_1 \sim V^{d_f}$ is observed, where $d_f$ denotes the volume fractal dimension. This result indicates that the percolation transition and associated FSS can still be well-defined in the regime $2<\lambda<3$ when data are sampled at the dynamic pseudocritical point, even though the subcritical phase is absent in the thermodynamic limit.

To estimate $d_f$, we fit $C_1$ using the FSS ansatz
\begin{equation}
C_1 = V^{d_f} (a_0 + a_1 V^{-\omega}). \label{eq-c1}
\end{equation}
The fitting results for different $\lambda$ and $\kappa$ are summarized in Table~\ref{t1}. We find that the fractal dimension $d_f$ depends on both $\lambda$ and $\kappa$, with smaller cutoffs leading to larger $d_f$. This suggests that suppressing hubs (large $\lambda$ or small $\kappa$) enhances connectivity among non-hub nodes, resulting in denser and more homogeneous cluster structures (large $d_f$). In addition, the mean-field value $d_f=2/3$ should be covered for $\kappa\to0$, which is well demonstrated in Fig.~\ref{f2}.

For percolation, the susceptibility-like quantity $\chi$ is expected to diverge at criticality as a scaling form of $\chi\sim V^{2d_f-1}$. However, previous studies~\cite{Cohen2002} have shown that $\chi$ vanishes at criticality for $2<\lambda<3$. As shown in Fig.~\ref{f3} (b), for a large cutoff ($\kappa = 1/2$), we indeed observe a power-law decay of $\chi$ with increasing system volume, indicating a suppression of critical fluctuations. This vanishing susceptibility can be attributed to fluctuation localization: in scale-free networks with $2<\lambda<3$ and large degree cutoffs, a small number of hubs dominate the connectivity. The giant cluster forms primarily via connections to these hubs, rather than through the merging of many intermediate-sized clusters. Consequently, critical fluctuations are localized around these hubs, and the extended fluctuations necessary for a divergent are suppressed, with this effect becoming more pronounced as the system size increases. In contrast, for smaller values of $\kappa$, the influence of hubs diminishes, and $\chi$ recovers the typical divergence expected in standard percolation transitions, see Fig.~\ref{f3} (b).

Furthermore, the lines in Fig.~\ref{f3} (b) represent the standard scaling form $\chi\sim V^{2d_f-1}$, using $d_f$ obtained from fitting $C_1$ (Table~\ref{t1}). The excellent agreement between the simulation results and the predicted scaling, even for $\kappa = 1/2$ where $d_f < 0.5$ and thus $2d_f - 1 < 0$, confirms the validity of the scaling relation between $C_1$ and $\chi$ across different $\kappa$ values.

\begin{figure}
\centering
\includegraphics[width=\columnwidth]{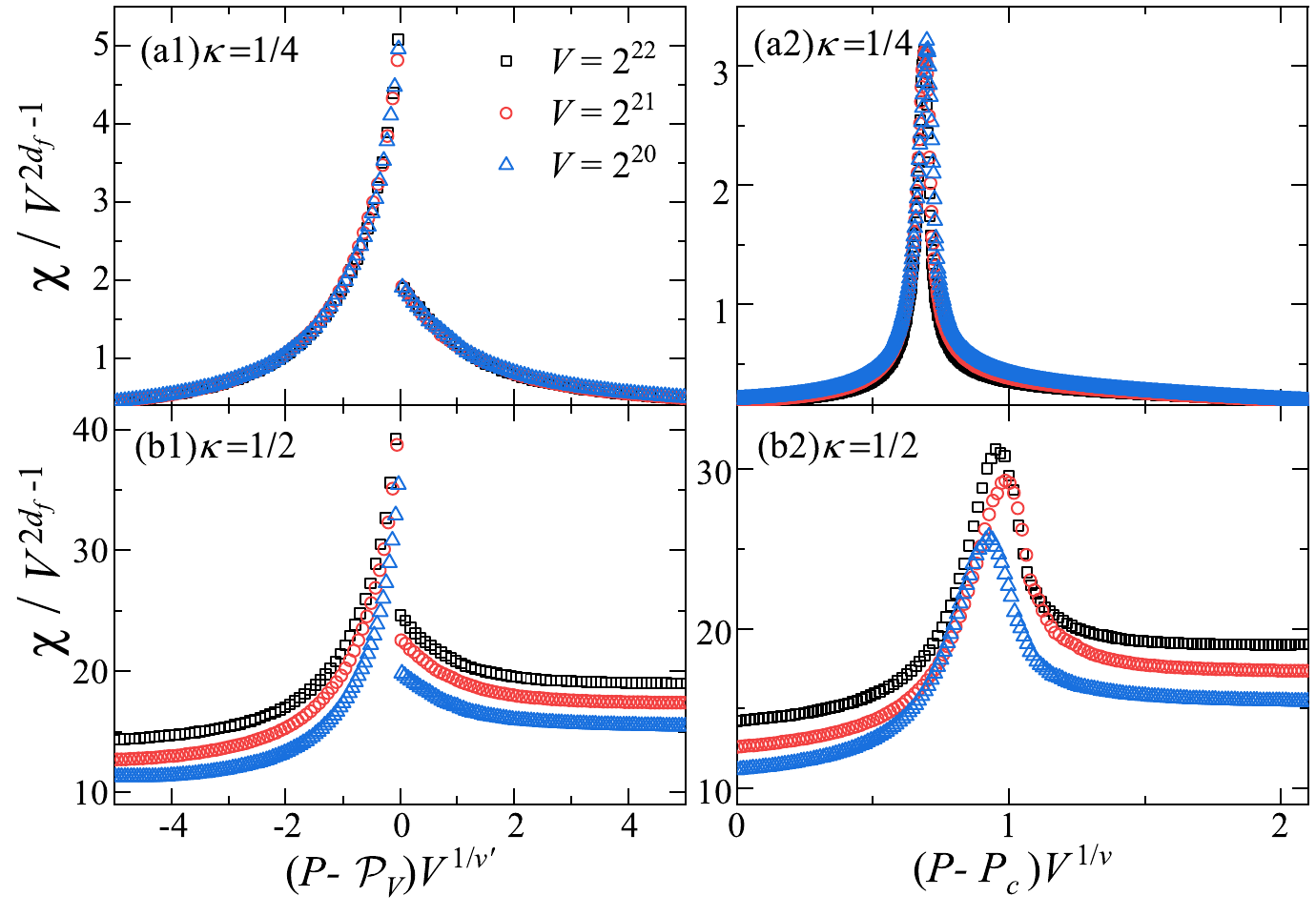}
\caption{(Color online) The FSS behavior of the susceptibility-like quantity $\chi$ near the dynamic pseudocritical point $\mathcal{P}_V$ and the critical point $P_c$ for $\lambda=2.5$. Panels (a1) and (a2) show $\chi/V^{2d_f-1}$ as a function of $(P-\mathcal{P}_V)V^{1/\nu'}$ and $(P - P_c)V^{1/\nu}$, respectively, for $\kappa=1/4$, where $\chi$ diverges. A better data collapse is observed in (a1), supporting the applicability of standard FSS theory around the pseudocritical point $\mathcal{P}_V$. Panels (b1) and (b2) show the same rescaled plots for $\kappa=1/2$, where $\chi$ vanishes in the thermodynamic limit. Data collapse cannot be achieved for both (b1) and (b2), suggesting an abnormal FSS. The exponents used here are taken from the fit results in Table~\ref{t1}. For (a1) and (b1), the discontinuity of $\chi$ at $\mathcal{P}_V$ comes from the event-based definition of $\mathcal{P}_V$.}
\label{f4}
\end{figure}

By applying standard FSS theory around $\mathcal{P}_V$, the behaviors of the susceptibility $\chi$ are represented as $\chi=V^{2d_f-1}\tilde{\chi}(x)$, where $x\equiv(P-\mathcal{P}_V)V^{1/\nu'}$ or $x\equiv(P-P_c)V^{1/\nu}$. To demonstrate this FSS behavior, we plot the rescaled susceptibility $\chi/V^{2d_f-1}$ as a function of $(P-\mathcal{P}_V) V^{1/\nu'}$ for $\kappa=1/4$ in Fig.~\ref{f4} (a1). By using the fitted values of $d_f$ and $1/\nu'$ from Table~\ref{t1}, the data for different system sizes collapse remarkably well near $\mathcal{P}_V$. The discontinuity of $\chi$ at $\mathcal{P}_V$ comes from the event-based definition of $\mathcal{P}_V$. In contrast, as shown in Fig.~\ref{f4} (a2), the same rescaling near $P_c$ using $x=(P-P_c)V^{1/\nu}$ cannot produce a satisfactory data collapse as that in Fig.~\ref{f4} (a1). These results not only supports the reliability of our extracted exponents $d_f$ and $\nu'$, but also indicates that the standard FSS theory is applicable around the dynamic pseudocritical point $\mathcal{P}_V$, with $\nu'$ playing the role of the correlation-length exponent instead of $\nu$.

For cases where the susceptibility vanishes in the thermodynamic limit (e.g., $\kappa = 1/2$), the rescaled susceptibility $\chi/V^{2d_f - 1}$ exhibits a converging height across different system sizes as $P \to \mathcal{P}_V^-$, consistent with the FSS behavior observed in Fig.~\ref{f3} (b). However, away from $\mathcal{P}_V$, a data collapse cannot be achieved, indicating that the scaling form holds only locally at $\mathcal{P}_V$. As shown in Fig.~\ref{f4} (b2), the rescaling with $x = (P - P_c)V^{1/\nu}$ also fails to produce a collapse in the whole range of $x$. Moreover, despite the theoretical result $P_c = 0$ for $2 < \lambda < 3$, the peak of $\chi$ in both Figs.~\ref{f4} (b1) and (b2) does not show an asymptotic shift toward $P_c = 0$ as system size increases. These findings suggest that while standard FSS holds near $\mathcal{P}_V$, the overall scaling behavior for $2 < \lambda < 3$ exhibits certain anomalies that cannot be fully captured by conventional FSS functions. This highlights the intrinsic complexity of FSS in systems with vanishing thresholds and strong structural heterogeneity.

\subsubsection{Cluster-number density}

\begin{figure}
\centering
\includegraphics[width=\columnwidth]{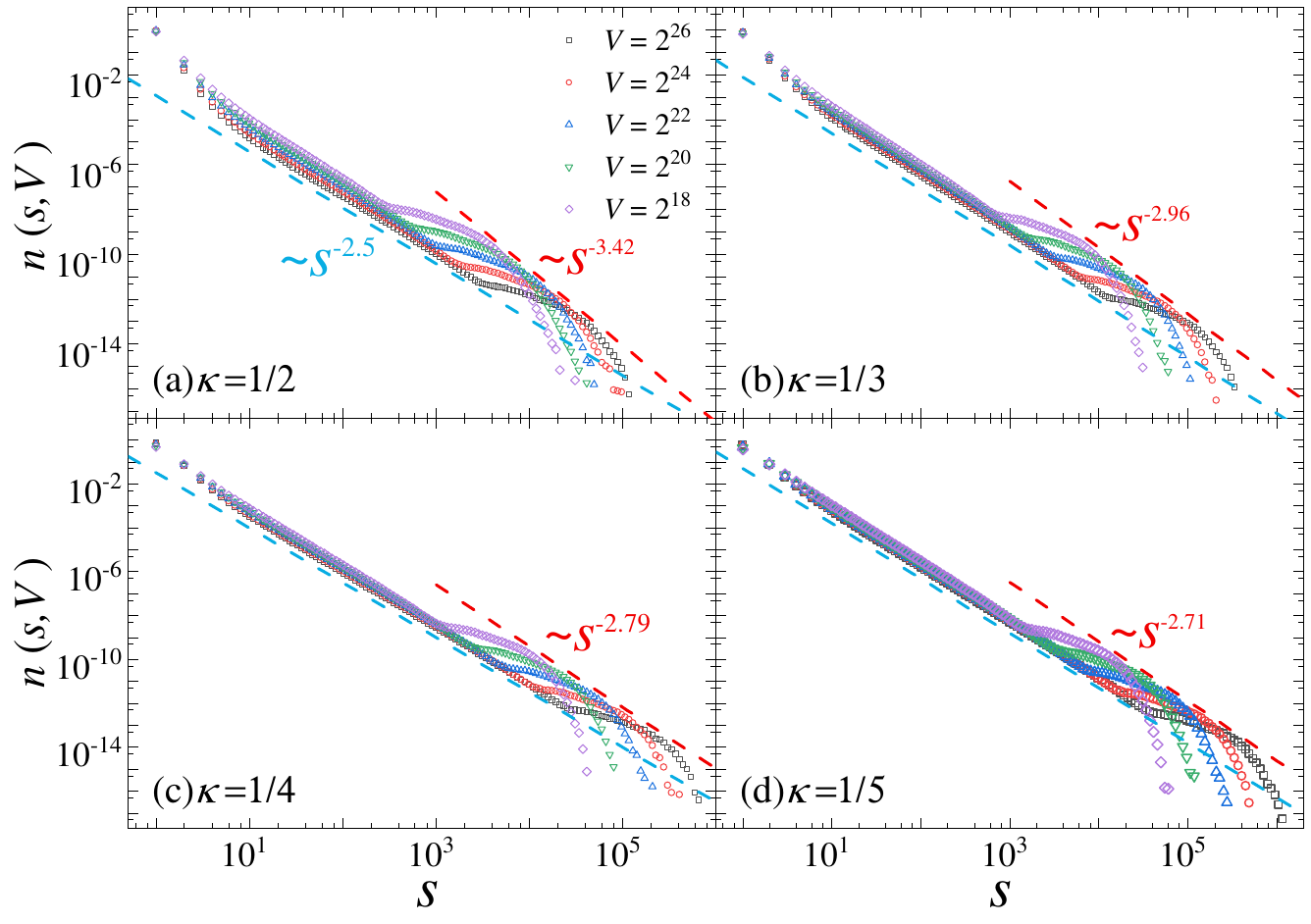}
\caption{(Color online) Cluster-number density $n(s, V)$ for $\lambda=2.5$ across different system volumes and degree cutoffs. Due to the vanishing percolation threshold in this regime, $n(s, V)$ decreases with increasing system size for all $s > 1$. Red dashed lines show power-law decay with exponent $\tau' = 1 + 1/d_f$, where the $\kappa$-dependent $d_f$ is obtained from the scaling of $C_1$ in Fig.~\ref{f3} (a). Blue dashed lines correspond to the Fisher exponent $\tau = \lambda$.}
\label{f5}
\end{figure}

At the critical point, percolation clusters are expected to be self-similar, with the cluster-number density following the scaling form $n(s,V) = s^{-\tau} \tilde{n}(s/V^{d_f})$, where hyperscaling predicts the Fisher exponent $\tau = 1 + 1/d_f$. However, for $2 < \lambda < 3$, the percolation threshold vanishes in the thermodynamic limit, and only isolated sites remain. Even at the dynamic pseudocritical point, the cluster-number density $n(s,V)$ for $s > 1$ decreases with increasing system size $V$, as demonstrated in Fig.~\ref{f5}.

Nevertheless, Fig.~\ref{f5} reveals a clear dependence of $n(s,V)$ on the degree cutoff. First, a power-law decay of $n(s,V)$ is still visible for large $s$ (red dashed lines), with an exponent consistent with the hyperscaling relation $\tau' = 1 + 1/d_f$, where $d_f$ is extracted from the scaling of $C_1$ in Fig.~\ref{f3} (a). Second, although the overall magnitude of $n(s,V)$ decreases for $V\to\infty$, a Fisher exponent $\tau$ can still be roughly identified (blue dashed lines). Remarkably, this exponent appears to be independent of the degree cutoff and aligns with the theoretical prediction $\tau = \lambda$ from the Potts model mapping~\cite{Lee2004} or the asymptote of the cluster-size distribution~\cite{Kryven2017}. This finding deviates from the earlier mean-field result $\tau = (2\lambda - 3)/(\lambda - 2)$~\cite{Cohen2002}, which gives $\tau = 4$ for $\lambda = 2.5$ -- clearly inconsistent with the numerical data in Fig.~\ref{f5}. This result is also clarified in Ref.~\cite{Cirigliano2024}. In fact, our fitted value of $\nu$ in Table~\ref{t1} agrees well with their analysis based on the generating-function approach, further supporting the inconsistency of the earlier mean-field prediction.

\subsection{$3<\lambda<4$}

\subsubsection{Dynamic pseudocritical points}

\begin{figure}
\centering
\includegraphics[width=\columnwidth]{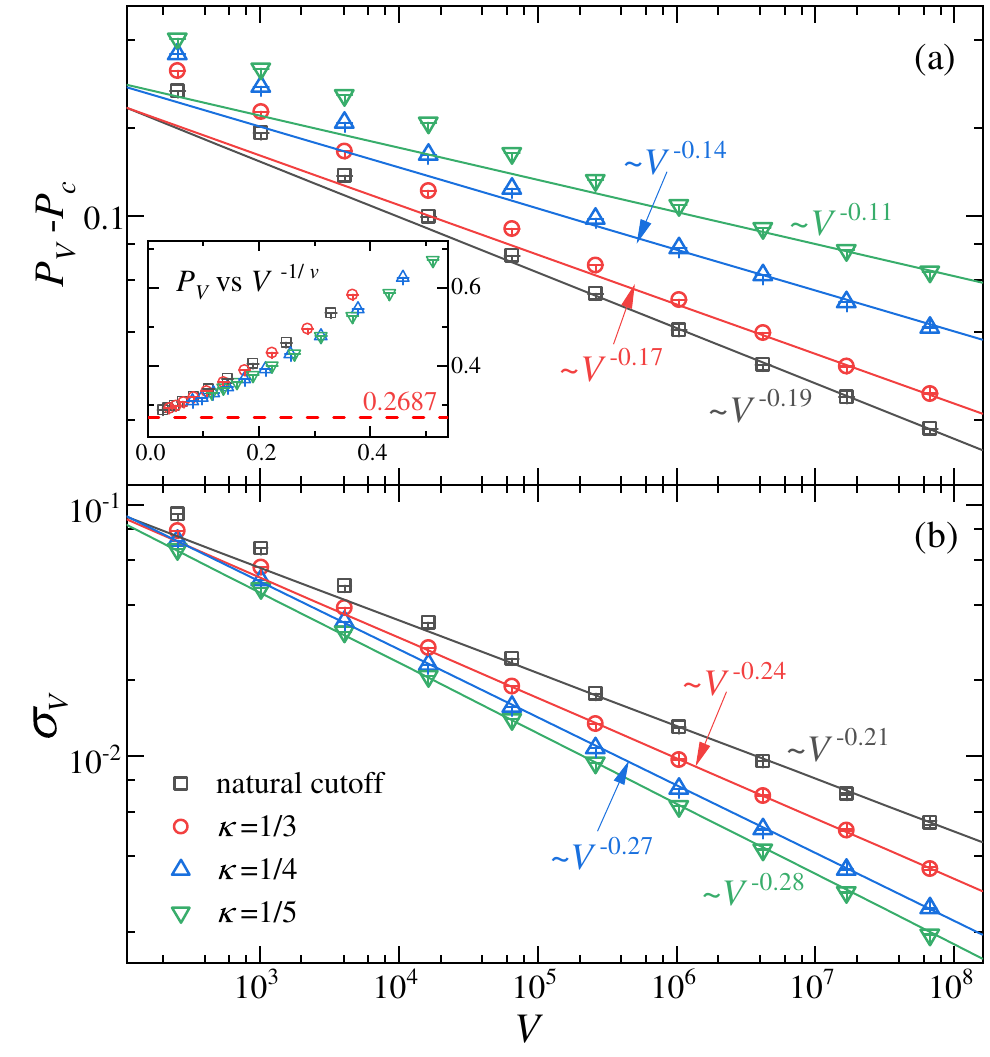}
\caption{(Color online) Asymptotic behavior of dynamic pseudocritical points for $\lambda=3.5$. (a) The distance $P_V - P_c$ between the critical point $P_c$ and the dynamic pseudocritical point $P_V$ is plotted against system volume $V$ for various degree cutoffs. The results exhibit clear power-law scaling $P_V-P_c\sim V^{-1/\nu}$ for large $V$, with the scaling exponent $1/\nu$ decreasing as the cutoff becomes smaller. The inset shows $P_V$ versus $V^{-1/\nu}$, where all curves extrapolate to the same $P_c \approx 0.2687$ as $V^{-1/\nu} \to 0$, indicating that the critical point is independent of the degree cutoff. (b) Fluctuation $\sigma_V$ of the dynamic pseudocritical point as a function of $V$. The FSS $\sigma_V \sim V^{-1/\nu'}$ yields a different exponent $1/\nu'$, which increases as $\kappa$ decreases.}
\label{f6}
\end{figure}

For $3 < \lambda < 4$, the Hurwitz zeta function $\zeta(s,a)$ in Eq.(\ref{eq-pcd}) is finite, yielding a nontrivial percolation threshold $P_c > 0$. Table~\ref{t1} lists the fitted values of $P_c$ for $\lambda = 3.3$, $3.5$, and $3.8$, obtained by fitting $P_V$ using the FSS ansatz in Eq.~(\ref{eq-fpv}). These values are consistent with the theoretical prediction from Eq.~(\ref{eq-pcd}). Notably, the percolation threshold given by Eq.~(\ref{eq-pcd}) is independent of the cutoff exponent $\kappa$, since $K \sim V^{\kappa}$ diverges for any $\kappa > 0$.

In contrast, the exponent $1/\nu$ governing the convergence $P_c - P_V \sim V^{-1/\nu}$ is $\kappa$-dependent, as illustrated in Fig.~\ref{f6} (a). The fitted values of $1/\nu$, also listed in Table~\ref{t1}, agree well with the theoretical expression $1/\nu = \kappa(\lambda - 3)$ from Eq.~(\ref{eq-pcv}), which reduces to $1/\nu = (\lambda - 3)/(\lambda - 1)$ under the natural cutoff~\cite{Cohen2002,Kalisky2006}. Plotting $P_V$ versus $V^{-1/\nu}$ in the inset of Fig.~\ref{f6} (a) confirms that all curves converge to the same critical point $P_c$ as $V^{-1/\nu} \to 0$, further validating the predicted exponent and percolation threshold.

A similar scaling behavior is observed for the fluctuation $\sigma_V \sim V^{-1/\nu'}$, as shown in Fig.~\ref{f6} (b) for $\lambda = 3.5$, with the fitted exponents $1/\nu'$ summarized in Table~\ref{t1}. Under the natural cutoff, we find $1/\nu' = 1/\nu = (\lambda - 3)/(\lambda - 1)$, in agreement with the theoretical prediction. Since $1/\nu' = (\lambda - 3)/(\lambda - 1)$ exceeds the mean-field value $1/3$ for $3 < \lambda < 4$, the exponent $1/\nu'$ decreases toward $1/3$ as the cutoff becomes more stringent (i.e., smaller $\kappa$), as illustrated in the inset of Fig.~\ref{f2}. This trend echoes the crossover behavior discussed earlier for the regime $2 < \lambda < 3$.

\subsubsection{Critical clusters}

\begin{figure}
\centering
\includegraphics[width=\columnwidth]{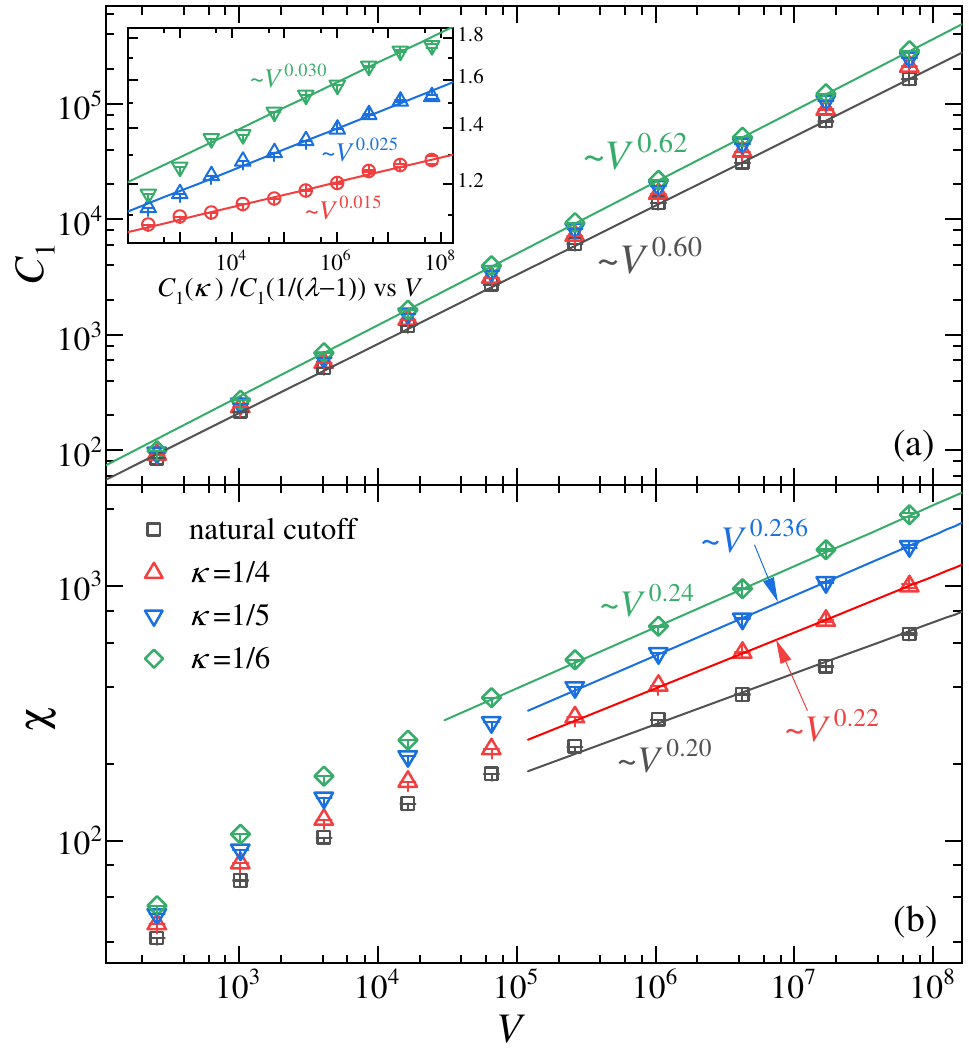}
\caption{(Color online) The FSS behaviors of the percolation on SF networks with $\lambda=3.5$. (a) The largest-cluster size $C_1$ as a function of system volume $V$ for various degree cutoffs. The lines indicate the scaling $C_1 \sim V^{d_f}$ for natural cutoff $\kappa=1/(\lambda=1)$ and $\kappa=1/6$, respectively, with fitted fractal dimensions $d_f$ listed in Table~\ref{t1}. The inset shows $C_1(\kappa)/C_1(1/(\lambda-1))$ for different $\kappa$. The nice scalings for large $V$ indicate that different $\kappa$ values lead to distinct fractal dimensions from that of natural cutoff. (b) $\chi$ as a function of $V$ for different degree cutoffs. The lines represent the scaling $\chi \sim V^{2d_f - 1}$, using $d_f$ from Table~\ref{t1}.}
\label{f7}
\end{figure}

According to the theoretical values of the critical exponents $\beta = 1/(\lambda - 3)$ and $1/\nu = (\lambda - 3)/(\lambda - 1)$ from Ref.~\cite{Cohen2002}, the volume fractal dimension of the largest cluster at criticality can be derived from the hyperscaling relation $d_f = 1 - \beta/\nu = (\lambda - 2)/(\lambda - 1)$, implying a $\lambda$-dependent fractal structure. For $\lambda = 3.3$, $3.5$, and $3.8$, this yields $d_f \approx 0.565$, $0.6$, and $0.643$, respectively, in good agreement with the fitted values listed in Table~\ref{t1}.

For a smaller degree cutoff, a clear power-law scaling $C_1 \sim V^{d_f}$ is still observed, as shown in Fig.~\ref{f7} (a) for $\lambda = 3.5$. The inset further demonstrates that the ratio $C_1(\kappa)/C_1(1/(\lambda - 1))$ grows as a power law when $\kappa < 1/(\lambda - 1)$, confirming that $d_f$ increases with decreasing $\kappa$. This indicates that the fractal dimension depends on both $\lambda$ and $\kappa$, and increases (i.e., becomes more compact) as hub nodes are more strongly suppressed, consistent with the trend observed for $2 < \lambda < 3$. In the limit $\kappa \to 0$, where the degree cutoff becomes very restrictive, the influence of the SF topology vanishes, and $d_f$ approaches the mean-field value $2/3$. This asymptotic behavior is clearly demonstrated in Fig.~\ref{f2}.

Moreover, the susceptibility-like quantity $\chi$ is found to scale as $\chi \sim V^{2d_f - 1}$, consistent with FSS theory for percolation. As shown in Fig.~\ref{f7} (b), simulation data for large systems follow this scaling law, using the $d_f$ values extracted from $C_1$, further supporting the internal consistency of the FSS framework for $3 < \lambda < 4$.

\begin{figure}
\centering
\includegraphics[width=\columnwidth]{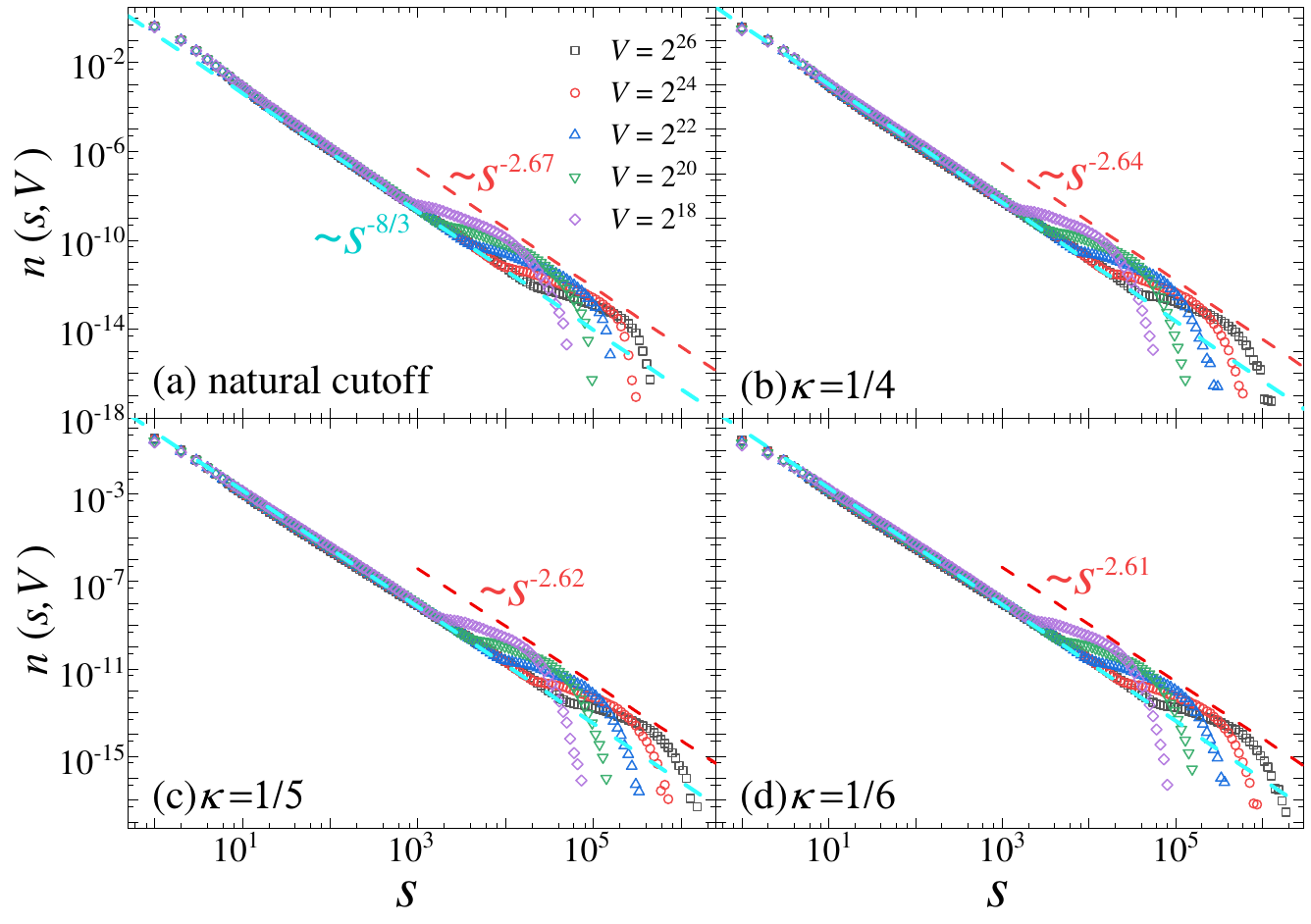}
\caption{(Color online) Cluster number density $n(s, V)$ for $\lambda=3.5$ across different system volumes and degree cutoffs. Blue dashed lines indicate the Fisher exponent $\tau=(2\lambda-3)/(\lambda-2)=8/3$. Red dashed lines show power-law decay with exponent $\tau'=1+1/d_f$, where the $\kappa$-dependent $d_f$ is obtained from the FSS of $C_1$ in Fig.~\ref{f7} (a). These results show that while the bulk of $n(s, V)$ is governed by $\lambda$-dependent scaling, the tail is also influenced by the degree cutoff.}
\label{f8}
\end{figure}

The $\kappa$-dependent fractal dimension also implies a $\kappa$-dependent Fisher exponent $\tau' = 1 + 1/d_f$. However, as in the regime $2 < \lambda < 3$, this exponent characterizes only the hump in the tail of the cluster-size distribution $n(s, V)$, as shown in Fig.~\ref{f8}. The bulk of $n(s, V)$ is instead governed by the theoretical value $\tau = (2\lambda - 3)/(\lambda - 2)$~\cite{Cohen2002,Lee2004}, which remains independent of the degree cutoff.

\subsection{$\lambda>4$}

\begin{figure}
\centering
\includegraphics[width=\columnwidth]{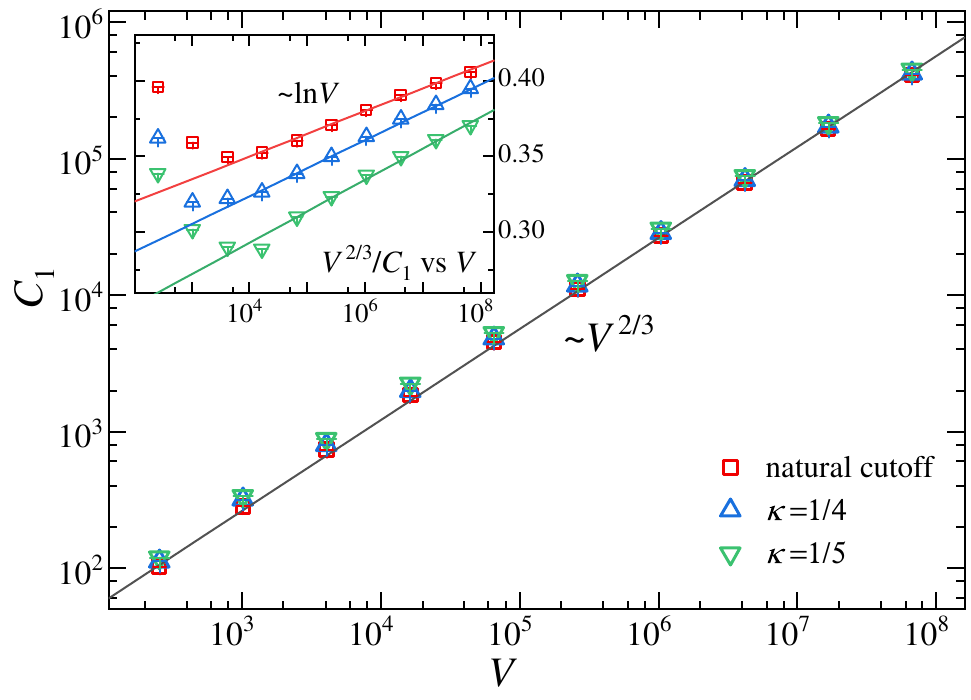}
\caption{(Color online) The FSS of $C_1\sim V^{d_f}$ across different degree cutoff for $\lambda=4.5$. The line shows the scaling of $C_1\sim V^{2/3}$. The inset plots the data of $V^{2/3}/C_1$ as a function of $V$ in a semi-logarithmic plot. For large $V$, the curve of $C_1/V^{2/3}$ approaches a straight line in the semi-logarithmic plot, suggesting a scaling behavior $V^{2/3}/C_1 \sim \ln V$, i.e., the FSS of $C_1$ contains a multiplicative logarithmic correction.}
\label{f9}
\end{figure}

For $\lambda > 4$, percolation on SF networks is expected to exhibit standard mean-field behavior, with $d_f = 2/3$ and $1/\nu = 1/\nu' = 1/3$. No crossover is expected in this regime. As shown in Table~\ref{t1}, for $\lambda = 6$, both the natural cutoff and a cutoff with $\kappa = 1/6$ yield $d_f$ and $1/\nu$ values consistent with the mean-field predictions within error bars.

As $\lambda$ approaches $4$, for instance at $\lambda = 4.5$, the fitted $d_f$ values from Eq.~(\ref{eq-c1}) are slightly below $2/3$, yielding $d_f \approx 0.64 \text{–} 0.65$. However, as illustrated in Fig.~\ref{f9}, the asymptotic scaling $C_1 \sim V^{2/3}$ appears to hold for large $V$, independent of $\kappa$. To examine this deviation, we plot $V^{2/3}/C_1$ versus $V$ on a semi-log scale (inset of Fig.~\ref{f9}). The data form an approximately straight line at large $V$, indicating $V^{2/3}/C_1 \sim \ln V$, and suggesting that logarithmic corrections accompany the scaling as $\lambda \to 4$.

To account for this, we incorporate a logarithmic term into the FSS ansatz
\begin{equation}
C_1 = V^{d_f} (\ln V + b)^X (a_0 + a_1 V^{-\omega}). \label{eq-c1ln}
\end{equation}
Allowing all parameters to vary freely yields unstable fits. Fixing $X = 1$ and testing various combinations of fixed $b$ and $\omega$, we obtain consistent estimates of $d_f$ across different $\kappa$ values, all converging to the mean-field value $2/3$ (Table~\ref{t1}). The reported error bars reflect the spread among these combinations.

In addition, Eq.~(\ref{eq-pcv}) suggests that when $\kappa < 1/3(\lambda - 3)$, before the mean-field value $\nu = 3$ is reached, $\nu$ may also depend on both $\kappa$ and $\lambda$, as noted in Ref.~\cite{Cirigliano2024}. However, strong finite-size effects, likely involving logarithmic corrections, prevent us from verifying this possibility with the present simulation accuracy.

\section{Discussion}  \label{sec-dis}

In this work, we systematically study the bond-percolation transitions on SF networks with tunable degree cutoffs. By analysing the FSS behaviors of the dynamic pseudocritical points $\mathcal{P}_V$, the largest-cluster size $C_1$, and the susceptibility-like quantity $\chi$, we demonstrated that the FSS behavior depends in a non-trivial way on both the degree exponent $\lambda$ and the cutoff exponent $\kappa$. Standard mean-field exponents reemerge whenever either $\lambda \ge 4$ or $\kappa\!\to 0$. The kernel of these crossovers is a structural change from heterogeneity to uniformity that underlines the pivotal role of hub nodes: suppressing hubs reduces network heterogeneity and drives the system towards the behavior of homogeneous or random graphs.

Based on extensive simulations, we summarize the FSS behavior as follows. For $\lambda > 4$, the system consistently exhibits mean-field behavior, regardless of $\lambda$ or $\kappa$. For $2<\lambda<4$, the fractal dimension $d_f$, the critical exponents $\nu$ (describing the shift of $\mathcal{P}_V$ from $P_c$), and $\nu'$ (describing the fluctuation of $\mathcal{P}_V$) all depend on $\kappa$ and deviate from their mean-field values. In contrast, except $\kappa=0$, for which the Fisher exponent takes the mean-filed value $\tau=5/2$, it is determined solely by $\lambda$, consistent with the relations $\tau = \lambda$ for $2 < \lambda < 3$ and $\tau = (2\lambda - 3)/(\lambda - 2)$ for $3 < \lambda < 4$. In the case of the natural cutoff, our results support $\nu=\nu'=(\lambda - 1)/|\lambda - 3|$, and $d_f$ is consistent with the hyperscaling relation $\tau = 1 + 1/d_f$. As $\kappa$ decreases, $\nu$ and $\nu'$ become distinct: $\nu'$ increases or decreases toward the mean-field value $\nu' = 3$, while $\nu$ increases above $3$ before shifting to $\nu = 3$ at $\kappa=0$. Meanwhile, the fractal dimension $d_f$ increases continuously toward the mean-field value $2/3$.

These crossover phenomena raise several open questions. First, in SF networks, intrinsic degree correlations may change alongside variations in the degree cutoff~\cite{Catanzaro2005,Lee2006}. Meanwhile, deliberately imposed correlations are known to alter percolation universality classes~\cite{Goltsev2008,Mizutaka2020,Wang2025}. Can the effects of a varying cutoff be understood as manifestations of intrinsic correlation changes, and might both be captured within a unified framework? Second, the degree distribution of empirical SF networks is often truncated in the form of an exponential decay~\cite{Clauset2009}, rather than by a hard cutoff. How such exponentially suppressed cutoffs influence the FSS behavior remains an open question. Third, although both $\nu$ and $\nu'$ vary with $\lambda$ and $\kappa$, our results consistently show $\nu > \nu'$, a relation also observed in high-dimensional percolation with open boundaries~\cite{Lu2024,Li2024a}. This resemblance suggests that both systems may share underlying organizational mechanisms yet to be uncovered. Fourth, our simulations reveal that SF networks exhibit strong finite-size corrections, often with mixed logarithmic contributions. These effects complicate the FSS analysis and call for more refined tools, such as the recently proposed gap method~\cite{Lu2024a}, which offers a promising way to capture more clean FSS. Overall, addressing these questions will require new theoretical insights, improved scaling arguments, and large-scale simulations that jointly consider degree heterogeneity, cutoff effects, and correlation structures.

\begin{acknowledgments}
The research was supported by the Fundamental Research Funds for the Central Universities (No.~JZ2023HGTB0220).
\end{acknowledgments}

\bibliography{ref}

\end{document}